\documentclass[aps]{revtex4}
\usepackage{amsthm,amsmath,amssymb,amscd,epsfig,float}
\def\ket#1{\left\vert #1 \right\rangle}
\def\bra#1{\left\langle #1 \right\vert}
\def\bkm#1#2{\left\langle #1 \vert #2 \right\rangle}
\def\ensavg#1{\left\langle #1 \right\rangle}

\def\u#1{\,\mathrm{#1}}

\def\trace#1{{\mathrm{Tr}\left[#1\right]}}

\def\RDM#1#2{{}^{#1}\!{#2}}

\def\refeqn#1{Eq.\ (\ref{Equation::#1})}

\def\refeqs#1#2{Eqs.\ (\ref{Equation::#1}) and (\ref{Equation::#2})}

\begin{document}
\title{Using tensor hypercontraction density fitting to achieve an $O(L^4)$ CISD algorithm}
\author{Neil Shenvi$^1$}
\author{Helen Van Aggelen$^{1,2}$}
\author{Weitao Yang$^1$}
\affiliation{$^1$ Dept. of Chemistry, Duke University, Durham, NC 27708}
\affiliation{$^2$ Dept. of Inorganic and Physical Chemistry, Ghent University, 9000 Ghent, Belgium}
\date{\today}

\begin{abstract}
Recently, Hohenstein et al\cite{Martinez:JCP12} introduced tensor hypercontraction density fitting to decompose the rank-4 electron repulsion 
integral tensor
as the product of five rank-2 tensors.  In this paper, we use this methodology to construct an algorithm which calculates the approximate ground state energy 
in $O(L^4)$ operations.
We test our method using several small molecules and show that we 
quickly approach the CISD limit with a small number of auxiliary functions.
\end{abstract}


\maketitle
The problem of the rapid growth of the electronic wavefunction with system size has plagued quantum chemistry for decades\cite{Levine:90,Szabo:96}.  There 
have been many attempts to 
conquer the `curse of dimensionality' and a large number of highly successful approximations have been developed.  
The difficulty inherent in electronic structure calculations is that correlated methods usually scale as high orders of 
the number of basis functions involved in the calculation.  A straightforward implementation of Hartree-Fock scales as $O(L^4)$ where $L$ is the number of basis 
functions.  Approximate methods employed to capture correlation usually face a trade-off between accuracy and 
efficiency\cite{Bartlett:ARPC81,HeadGordon:JPC96}.  Perturbative treatments such as MP2 and MP4 scale as $O(L^5)$ and $O(L^6)$, 
respectively\cite{Bartlett:ARPC81}.  Configuration interaction methods, which add in single-, double-, triple- and quadruple order excitations, also form a 
hierarchy of methods which scale as $O(L^5)$ and higher\cite{Bartlett:RMP07}.  A plethora of other, highly accurate 
methods based on coupled-cluster theory\cite{Bartlett:RMP07}, 2-particle reduced density 
matrices\cite{Mazziotti:PRA98,Mazziotti:JCP04}, and reduced active 
space diagonalization\cite{Roos:CPL89} also scale as at least $O(L^6)$.  Due to 
this high scaling, these methods are generally not applicable beyond small molecules.  Instead, quantum chemists have increasingly turned to density functional 
theory (DFT), which offers low $O(L^4)$ or even $O(L^3)$ formal scaling while managing to capture varying amounts of the electronic 
correlation\cite{Yang:CR12}.  
Nonetheless, the 
search for efficient wavefunction-based methods that are competitive with DFT has continued, motivated by the importance of strong correlation 
in many systems such as transition metal clusters, solid state devices, and molecules far from their equilibrium geometries.

One approach which attemps to reduce both the cost and scaling of correlated electronic structure methods is the decomposition of the 
electronic repulsion integral (ERI) tensor, which is naturally a rank-4 object, into products of lower-rank objects.  
Resolution-of-the-identity techniques\cite{Feyereisen:CPL93,Ahlrichs:CPL95,Fruchtl:TCA97,Haser:TCA97} are one subset of this approach, 
as are pseudospectral 
methods\cite{Martinez:JCP93,Martinez:JCP94,Martinez:JCP95}.  
Our algorithm was strongly motivated by the recent work of Hohenstein et al.\cite{Martinez:JCP12}, who developed a method
which they named `tensor hypercontraction density fitting' (THC-DF).  
The authors showed that their decomposition could be used to derive $O(L^4)$ MP2 and MP3 algorithms and 
suggested that it could also be used to improve the efficiency of a wide variety of electronic structure methods.  
Here, we apply their 
idea to the CISD method and show that it yields a dramatic
reduction of computational cost, from $O(L^6)$ for normal CISD to $O(L^4)$ for our method.  Our method will rigorously approach the 
traditional CISD energy as the number of auxiliary functions is increased.  Therefore, our method will face the same challenges, such as lack 
of size-extensivty, as the traditional CISD method.  However, we believe that the substantial reduction in computational scaling exhibited
by our method more than makes up for the well-known shortcomings of CISD, particularly given the fact that numerous methods exist for the 
correction of these shortcomings\cite{Dierksen:JCP94}.  

The electronic Hamiltonian of an atom or molecule can be written in 2nd-quantized notation as
\begin{eqnarray}
\hat{H} &=& \hat{H}_1 + \hat{H}_2 \\
&=&\sum_{ik}{h^i_k \hat{c}^\dagger_i \hat{c}_k} + \sum_{ijkl}{\epsilon^{ik}_{jl} \hat{c}^\dagger_i \hat{c}_k \hat{c}^\dagger_j \hat{c}_l }
\end{eqnarray}
where $h^i_k$ is the rank-2 matrix of 1-electron integrals and $\epsilon^{ij}_{kl} = (ik|jl)$ is the rank-4 electronic repulsion integral tensor. 
The ERI can be decomposed using a set of $P_H$ auxiliary functions as
\begin{equation} 
\label{Equation::HDFDef}
\epsilon^{ik}_{jl} = \sum_{a,b=1}^{P_H}{x_{ia}x_{ka}Z_{ab}x_{jb}x_{lb}}.
\end{equation}
This decomposition can be achieved trivially by writing $\epsilon$ as a $L^2\times L^2$ matrix and diagonalizing it to obtain eigenvector $v^\lambda_{ik}$ and 
eigenvalues $d^\lambda$.
We can then pick some arbitrary orthonormal 1-electron basis $x_{ia}$ and decompose the eigenvectors $v^\lambda_{ik}$ into a product of rank-1 objects by 
solving $L^2$ linear equations
\begin{equation}
v^\lambda_{ik} = \sum_{a=1}^{L^2} y^\lambda_a x_{ia} x_{ka}.
\end{equation}
Letting 
\begin{equation} \label{Equation::EpsExact}
Z_{ab} = \sum_{\lambda=1}^{L^2}{y^\lambda_a d^\lambda y^\lambda_b}
\end{equation}
we obtain the exact decomposition in \refeqn{HDFDef}.  The problem with this decomposition is that 1) it involves a large number of auxiliary basis functions 
$P_H=L^2$ and 2) it is based on the diagonalization of the $L^2 \times L^2$ matrix $\epsilon$, which requires $O(L^6)$ operations.

One of the most important conclusions of \cite{Martinez:JCP12} was that real molecular ERI tensors can be well-approximated to $\u{mH}$ accuracy by the 
expression in \refeqn{HDFDef} with a number of auxiliary functions that scale only as $P_H=O(L)$.  The authors also devised an algorithm to obtain the 
decomposition in \refeqn{HDFDef} in $O(L^4)$ operations.  
Although this previous result is crucial to ensuring the low scaling of our overall algorithm, for the purposes of testing the intrinsic accuracy 
of our method, we will use the exact decomposition prescribed in \refeqn{EpsExact} to avoid any errors inherent to the approximation of the ERI tensor.  
Another important feature of the THC-DF decomposition is that it enables the efficient $O(L^3)$ transformation of the ERI from the atomic orbital 
basis to any other arbitary one-electron basis, such as the orthonormal molecular orbital basis.  
Because of this simple transformation property, we will assume that the ERI is already written an orthonormal basis such as the molecular orbital basis.
 
Our CISD method itself is based on a parametrization of a single-Slater determinant reference wavefunction, usually taken to be the 
Hartree-Fock ground state.  We parametrize our trial 
wavefunction $\ket{\phi}$ as follows:
\begin{equation} \label{Equation::PhiDef}
\ket{\phi} = \hat{A}\ket{\Psi}, \hat{A} = A_0+\sum_{ikjl}{A^{ik}_{jl} \hat{c}^\dagger_i \hat{c}_k \hat{c}^\dagger_j \hat{c}_l}
\end{equation}
where $\ket{\Psi}$ is our single-Slater determinant reference wavefunction and $\hat{A}$ is an arbitrary 2-body excitation operator {\it which need not be 
expressed in the natural orbital basis of $\ket{\Psi}$}.  This latter condition is important for understanding our algorithm, since $\hat{A}$ need not excite 
only from occupied to virtual orbitals as is often assumed in CISD or CCSD 
derivations.  The constant $A_0$ is added to $\hat{A}$ to guarantee that the original reference function is contained within our ansatz.  
Because of the spin-invariance of non-relativistic molecular Hamiltonians, we can constrain our ansatz to preserve the number of $\alpha$ and $\beta$ 
electrons.  For this reason, our actual ansatz is
\begin{equation}
\hat{A} = A_0 + \hat{A}_{\alpha\alpha}+\hat{A}_{\alpha\beta}+\hat{A}_{\beta\beta}.
\end{equation}
where the operators $\hat{A}_{\alpha\alpha},\hat{A}_{\alpha\beta}$ and $\hat{A}_{\beta\beta}$ act on two spin-up, one spin-up and one spin-down, or two 
spin-down electrons, respectively.  In what follows, we will ignore this simplification, as it needlessly complicates notation.

Noting that $A^{ik}_{jl}$ is a rank-4 tensor, we apply 
the THC-DF decomposition from \cite{Martinez:JCP12} and express it as
\begin{equation} \label{Equation::AHDF}
A^{ik}_{jl} = \sum_{a,b=1}^{P_A}{\chi_{ia}\chi_{ka}{\mathcal Z}_{ab}\chi_{jb}\chi_{lb}}
\end{equation}
If we let $P_A = L^2$, we can represent any 2-body excitation operator $\hat{A}$ and hence any CISD wavefunction $\ket{\phi}$.  For smaller values of $P_A$, we do not span the 
entire CISD space but nonetheless have a very efficient and interesting representation of our trial wavefunction.  As we will show in our examples, we are able to obtain excellent approximations to 
the CISD energy with $P_A < L$.

To obtain an approximate ground state energy, we substitute our trial wavefunction into the standard variational energy expression to obtain
\begin{eqnarray} \label{Equation::EPhiDef}
E_\Phi &=& \bra{\Phi}\hat{H}\ket{\Phi}/\bkm{\Phi}{\Phi} \\
\label{Equation::EPhiDef2}
&=& \left(\ensavg{\hat{A}\hat{H}_1\hat{A}}+\ensavg{\hat{A}\hat{H}_2\hat{A}}\right)/\ensavg{\hat{A}\hat{A}}
\end{eqnarray}
where expectation values are taken over the reference wavefunction $\ket{\Psi}$.  The expressions on the right-hand-side of \refeqn{EPhiDef2} look formidable 
because they involve calculating the 
expectation value of 4- to 6-body operators.  It is at this stage that the THC-DF decomposition becomes crucially important because it allows us to evaluate the 
quantities in \refeqn{EPhiDef2} with 
only $O(L^4)$ operations.  To see how, let us consider the most complicated term in \refeqn{EPhiDef2}, $\ensavg{\hat{A}\hat{H}_2\hat{A}}$, which involves the 
evaluation of a 6-body operator.  Plugging in our decompositions in 
\refeqs{HDFDef}{AHDF}, we obtain
\begin{equation} \label{Equation::AH2ADef}
\ensavg{\hat{A}\hat{H}_2\hat{A}} = 
\sum{\left(\chi_{ia}\chi_{ka}{\mathcal Z}\chi_{jb}\chi_{lb}\right)\left(x_{mc}x_{oc}Z_{cd}x_{nd} x_{pd}\right)\left(\chi_{qe}\chi_{se}{\mathcal Z}_{ef}
\chi_{rf}\chi_{tf}\right)\ensavg{
\hat{c}^\dagger_i \hat{c}_k\hat{c}^\dagger_j \hat{c}_l\hat{c}^\dagger_m \hat{c}_o\hat{c}^\dagger_n \hat{c}_p\hat{c}^\dagger_q \hat{c}_s\hat{c}^\dagger_r 
\hat{c}_t}}
\end{equation}
where we sum over all repeated indices.  

Using the canonical fermionic anticommutation relations, we can express the operator in \refeqn{AH2ADef} in 
canonical ordering, with all creation operators on the left and all annihilation operators on the right.  Once the operators are canonically ordered, their 
expectation values over the single Slater determinant wavefunction $\ket{\Psi}$ can
be written exactly as antisymmetrized products of the 1-particle reduced density matrix\cite{Mazziotti:IJQC98,Mazziotti:CPL04},
\begin{equation}
\RDM{1}{D}_{ik} = \bra{\Psi}\hat{c}^\dagger_i \hat{c}_k\ket{\Psi}.
\end{equation}
For instance, we can write
\begin{equation}
\ensavg{\hat{c}^\dagger_i \hat{c}^\dagger_j \hat{c}_l \hat{c}_k} = \RDM{1}{D}^i_k\RDM{1}{D}^j_l-\RDM{1}{D}^i_l\RDM{1}{D}^j_k.
\end{equation}
Expressing the expectation value of an $n$-body operator over the wavefunction $\ket{\Psi}$ will require $n!$ separate terms, so the result will be
lengthy.

The question that immediately arises is whether these lengthy expressions can be evaluated efficiently.  At this point, the THC-DF decompositions employed in 
\refeqs{HDFDef}{AHDF} 
become extremely important.  Because both the Hamiltonian and excitation operators are decomposed in terms of rank-2 tensors, a given index can be connected 
to \emph{at most} three other indices. For instance, the index $a$ in \refeqn{AH2ADef} is involved only in the terms $\chi_{ai}, \chi_{ak}$ and ${\mathcal 
Z}_{ab}$.  
As a result, we can hold $i,k$ and $b$ constant and perform the sum over $a$, storing the result
\begin{equation}
X_{ikb} = \sum_{a}{\chi_{ai}\chi_{ak}{\mathcal Z}_{ab}}
\end{equation}
for future use.  When we implement all the terms in \refeqn{AH2ADef}, we find that we can evaluate all terms in our expression by summing over a single index 
while holding at most three others constants.  
Provided that both $P_H$ and $P_A$ scale as $O(L)$, this fact entails that the expected energy can be evaluated in $O(L^4)$ operations, albeit with a large 
prefactor.  Furthermore, when we take the analytic gradient of our energy with respect to the variational parameters $\chi_{ai}$ and 
${\mathcal Z}_{ab}$, we find that this result also requires only $O(L^4)$ operations.  An illustrative example of the terms involved in our calculation is 
provided in the Appendix.

Our final algorithm is implemented as follows:  We begin with some initial guess of the variational parameters $\chi_{ia}$ and ${\mathcal Z}_{ab}$.  
Currently, we have 
found that selecting a random matrix $\chi_{ia}$ and setting ${\mathcal Z} = 0$ appears to work well, especially when $P_A$ is sufficiently large.  For small 
values of $P_A < 5$, the minimization can become trapped in local minima, but sampling a small number of initial conditions guarantees that a repeatable 
global minimum will be found.  We then perform a quasi-Newton minimization on the expected energy $E_\Phi$ in \refeqn{EPhiDef}, seeking to minimize it with 
respect to our variational parameters $\chi_{ia}$ and ${\mathcal Z}_{ab}$ using the LBFGS optimization algorithm described in \cite{Nocedal:89, Nocedal:97}.  
The minimiation converged to within $1 \u{mH}$ of its final result within $300$ iterations for all the molecules studied and we expect that miminization time 
can be greatly reduced by using better initial guesses and better approximations to the Hessian in the LBFGS algorithm.  Nonetheless, since both the 
calculation of the energy and its gradient require $O(L^4)$ operations, our minimization algorithm can likewise be accomplished in $O(L^4)$ time. Upon 
minimization, the energy we obtain will be an upper bound on the CISD energy and will approach the CISD result as the number of auxiliary functions $P_A$ is 
increased.

Our first set of results in Table I shows the electronic energies obtained for a selection of small molecules in a minimal STO-6G basis set.  The RHF and CISD 
results and the ERIs are calculated using Gaussian09\cite{Gaussian:09}.  The remainder of the table shows the energies calculated by our method as a function 
of the size of the auxiliary basis $P_A$.  We see that our method converges to the CISD result as we increase the size of the auxiliary basis.  More surprising 
is how few auxiliary functions are 
needed to obtain a very accurate approximation to the CISD result.  For instance, almost 95\% of the CISD correlation energy in the molecule LiH could be 
recovered by our algorithm with only \emph{two} auxiliary functions $P_A=2$.  For other molecules a larger number of auxiliary functions were required to 
ensure good convergence to the CISD answer.  But in all cases, in order to converge recover 98\% of the correlation energy present in the exact CISD result, no 
more than $P_A = 6$ auxiliary functions were required.  This result is important because $P_A$ must scale as $O(L)$ to ensure that our overall algorithm retains 
a scaling of $O(L^4)$.  Note that for large $P_A$, our algorithm recovers slightly \emph{more} correlation energy than is recovered by the CISD algorithm 
due to spin-contamination in our algorithm.  While the CISD algorithm is spin-projected such that only singlet states are included, our algorithm allows 
mixing into states of other spin-multiplicities, yielding energies that are ever so slightly lower than the spin-pure result.  In all cases, this effect 
was negligible, amounting to less than $0.3 \u{mH}$.

\begin{table}
\begin{tabular}{|l||c|c|c|c|c|c|}
\hline
 & $E_g$ (H) & \multicolumn{5}{|c|}{Correlation Energy (mH)} \\
\hline
Molecule & $E_{HF}$ & CISD & $P_A = 2$ & $P_A = 4$ & $P_A = 6$ & $P_A = 10$ \\
\hline
\hline
BH & -27.1484 & 55.7 & 37.8 & 55.4 & 55.8 & 55.9 \\
\hline
LiH & -8.9182 & 20.9 & 20.1 & 21.0 & 21.0 & 21.1 \\
\hline
BeH2 & -19.0803 & 34.5 & 29.7 & 32.2 & 34.7 & 34.8 \\
\hline
CH2 & -44.7828 & 58.1 & 35.7 & 55.0 & 57.2 & 58.2 \\
\hline
HF & -103.1456 & 66.7 & 65.0 & 66.1 & 66.4 & 66.7 \\
\hline
H2O & -84.7683 & 50.7 & 46.0 & 47.2 & 50.0 & 50.5 \\
\hline
\end{tabular}
\caption{Ground state energies for a variety of small molecules in the STO-6G basis set.  As the number of auxiliary 
functions, $P_A$, is increased, the energy obtained by our algorithm approach the exact CISD result.  In all cases, our method with $P_A 
= 6$ recovers more than $98\%$ of the CISD correlation energy.
}
\end{table}

\begin{table}
\begin{tabular}{|l||c|c|c|c|c|c|}
\hline
 & $E_g$ (H) & \multicolumn{5}{|c|}{Correlation Energy (mH)} \\
\hline
Molecule & $E_{HF}$ & CISD & $P_A = 2$ & $P_A = 4$ & $P_A = 6$ & $P_A = 10$ \\
\hline
\hline
BH  & -27.2559 & 59.9 & 32.1 & 52.8 & 59.0 & 59.8 \\
\hline
LiH  & -8.9475 & 19.0 & 17.5 & 18.6 & 19.1 & 19.2 \\
\hline
BeH2 & -19.1161 & 39.4 & 16.8 & 34.4 & 38.8 & 39.3 \\
\hline
CH2 & -44.8861 & 84.3 & 36.6 & 58.2 & 77.2 & 81.6 \\
\hline
HF & -103.6346 & 144.9 & 67.8 & 128.1 & 139.6 & 143.3 \\
\hline
H2O & -85.0717 & 130.2 & 47.3 & 111.0 & 121.0 & 125.8 \\
\hline
\end{tabular}
\caption{Ground state energies for a variety of small molecules in the 6-31G basis set.  As the number of auxiliary 
functions, $P_A$, is increased, the energy obtained by our algorithm approach the exact CISD result.  In all cases, our method with $P_A = 10$ recovers more 
than $96\%$ of the CISD correlation energy.}
\end{table}

Table II shows the same results for the same molecules using the larger 6-31G basis.   Again, we see a rapid convergence to the exact CISD result as a 
function of $P_A$.  These results also support the inference that only $P_A = O(L)$ auxiliary functions are needed to converge to the exact result, since in 
all cases studied, setting $P_A = L$ recovers $>97\%$ of the exact CISD correlation energy.

Finally, Figure 1 offers further support for our contention that our algorithm scales as $O(L^4)$.  This plot shows the computational cost of calculating 
the energy and gradient of a system of $N$ non-interacting H$_2$ molecules as a function of $N$.  Because the molecules are strictly 
non-interacting, $P'_H = N 
P_H$, where $P'_H$ and $P_H$ 
are the number of auxiliary functions needed to describe the composite and single system, respectively.  Similarly, we use $P'_A = N P_A$ auxiliary functions to 
represent our excitation operator.
As a result, this system provides the perfect test case for measuring the computational scaling of 
out algorithm.  The result in Fig. 1 shows that our algorithm scales as $O(L^4)$.  This measurement is consistent with an examination of our code.  The 
slow step in our algorithm is a series of nested do-loops that sum four indices over the range $1\ldots P_A$ or $1\ldots P_H$.  Provided that $P_A$ and $P_H$ 
both scale as $O(L)$, our algorithm then has a complexity of $O(L^4)$, as we observe.

\begin{figure}
\includegraphics{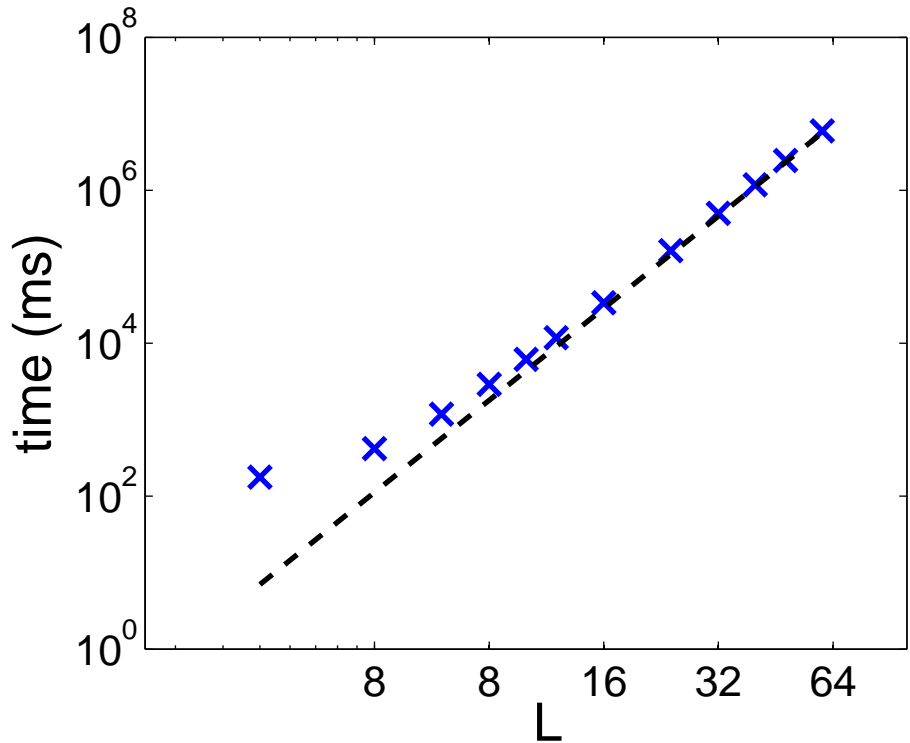}
\caption{The computational cost of each energy/gradient calculation as a function of the number of basis function $L$ for a system of $N$ noninteracting H$_2$ 
molecules in the STO-6G basis.  The blue Xs show the numerical results; for comparison, the dashed line shows $L^4$ scaling.  As expected, our algorithm scales 
asymptotically as $O(L^4)$
}
\end{figure}

\

There are numerous areas of future research.  One of the most important issues is performing the efficient THC-DF decomposition of the Hamiltonian.  Although 
the authors of \cite{Martinez:JCP12} introduced a method to decompose a molecular Hamiltonian in terms of $P_H = O(L)$ auxiliary functions using $O(L^4)$ 
operations, they recognized that there were very large prefactors involved in this process.  For this reason, it is worth investigating whether we can 
construct such a decomposition more efficiently, even if the construction process retains the same asymptotic scaling.

A related issue is the prefactor associated with the electronic structure algorithm presented in this paper.  The slow step of our algorithm is the 
calculation 
of the energy and gradient of the 6-body term $\ensavg{\hat{A}\hat{H}_2\hat{A}}$.  
Since this expectation value involves six creation and six 
annihilation operators, there are $6! = 720$ possible matchings of the operators, each of which contributes a unique term to the energy and gradient.  
The code required to perform such calculations is immense; our final program contained over 500,000 lines of FORTRAN code even when effort was made to store 
intermediate results and eliminate redundancy.  Major improvement could be made to our algorithm either by developing analytic methods to reduce the cost of 
energy/gradient evaluation or by developing methods which identify and approximate negligible terms in our equations.

A final major area of research is the extension of our method to coupled-cluster algorithms.  
One of the key elements of coupled cluster theory is that the excitation operator $T_2$ must only connect occupied orbitals to virtual 
orbitals; there can be no non-zero elements which act within the occupied or virtual subspaces or which connect virtual orbitals to occupied orbitals. 
However, if we make this assumption, it is not 
clear that the excitation operator $T_2$ can still be efficiently written in the THC-DF form, which is vital to the efficiency of our algorithm.   In other 
words, 
the two approximations which are crucial to an THC-DF-based coupled-cluster theory may be mutually incompatible.  Future studies will have to assess whether 
this is indeed the case or whether either theory can be modified to produce an efficient coupled-cluster algorithm.  
It is also interesting that it has already 
been shown in a different context that generalized coupled-cluster methods, which do not make the occupied/unoccupied distinction, can exactly represent 
the ground state many-body wavefunction\cite{Nooijen:PRL00}.  Our results provide additional motivation for investigating generalized coupled-cluster methods.

In conclusion, we have described an $O(L^4)$ electronic structure algorithm which can be tuned in its accuracy through the value of the parameter $P_A$.  When 
this parameter is $0$, we recover the Hartree-Fock result.  When the parameter is $O(L)$, we rapidly approach the exact CISD result.  Although
the percentage of correlation energy recovered by the CISD approach decreases to $0$ as a function of the system size due to its lack of size extensivity, the 
many known approaches to recovering size-extensivity within CISD should be applicable to our method, thereby resotring its utility even for large 
systems\cite{Dierksen:JCP94}.  Our hope is that the 
efficiency of this algorithm will make it computationally competitive with existing low-scaling methods like Hartree-Fock and density functional theory, 
bringing explicitly correlated wavefunction-based methods into the realm of practical calculation even for large systems.


\begin{acknowledgments}
NS and WY would like to acknowledge support from the National Science Foundation (CHE-09-11119).  HvA thanks the FWO-Flanders and Duke University for support.
\end{acknowledgments}

\section{Appendix}
The total variational energy is given by
\begin{equation}
E_\Phi = \left(\ensavg{\hat{A}\hat{H}_1\hat{A}}+\ensavg{\hat{A}\hat{H}_2\hat{A}}\right)/\ensavg{\hat{A}\hat{A}}
\end{equation}
where the $\hat{A}$ operator is given by
\begin{equation}
\hat{A} = A_0+\hat{A}_{\alpha\alpha}+\hat{A}_{\alpha\beta}+\hat{A}_{\beta\beta}
\end{equation}
and where the operators $\hat{H}_1$ and $\hat{H}_2$ conserve spin such that they can be written as
\begin{eqnarray}
\hat{H}_1 &=& \hat{H}_1^\alpha + \hat{H}_1^\beta \\
\hat{H}_2 &=& \hat{H}_2^{\alpha\alpha} + \hat{H}_2^{\alpha\beta} + \hat{H}_2^{\beta\beta}
\end{eqnarray}
For our demonstration, we will consider only the contributions to the total energy that come from the term
\begin{equation}
E = \ensavg{A_0 \hat{H}_1^\alpha \hat{A}_{\alpha\alpha}}
\end{equation}

Using the THC-DF decomposition for the excitation operator, we obtain the expression
\begin{equation} \label{Equation::EAHB}
E = A_0 \sum_{ikabjlmn}{ h_{ik} \chi_{aj}\chi_{al} {\mathcal Z}_{ab} \chi_{bm} \chi_{bn} 
\ensavg{\hat{c}^\dagger_i \hat{c}_k \hat{c}^\dagger_j \hat{c}_l \hat{c}^\dagger_m \hat{c}_n} }
\end{equation}
To evaluate the expectation value in \refeqn{EAHB}, we use the canonical fermionic anticommutation relations.  For convenience, we define the 1-hole reduced 
density matrix $\RDM{1}{Q}^i_k = \delta^i_k - \RDM{1}{D}^i_k$.  After some tedious algebra, we obtain
\begin{eqnarray} \label{Equation::cikjlmn}
\ensavg{\hat{c}^\dagger_i \hat{c}_k \hat{c}^\dagger_j \hat{c}_l \hat{c}^\dagger_m \hat{c}_n} &=& 
\RDM{1}{D}^i_k \RDM{1}{D}^j_l \RDM{1}{D}^m_n
+\RDM{1}{D}^i_k \RDM{1}{D}^j_n \RDM{1}{Q}^l_m
+\RDM{1}{D}^i_l \RDM{1}{Q}^k_j \RDM{1}{D}^m_n\\
\notag &&+\RDM{1}{D}^i_n \RDM{1}{Q}^k_m \RDM{1}{D}^j_l
-\RDM{1}{D}^i_l \RDM{1}{Q}^k_m \RDM{1}{D}^j_n
+\RDM{1}{D}^i_n \RDM{1}{Q}^k_j \RDM{1}{Q}^l_m
\end{eqnarray}
If we plug \refeqn{cikjlmn} into \refeqn{EAHB}, we obtain
\begin{eqnarray}
\label{Equation::EAHBexpand}
E &=&  A_0\sum_{ab}\Big(\trace{h\cdot D}\left[\chi\cdot D\cdot \chi^\dagger\right]_{aa} {\mathcal Z}_{ab} \left[\chi\cdot D\cdot \chi^\dagger\right]_{bb} \\
\notag
& & +\trace{h\cdot D}\left[\chi\cdot D\cdot \chi^\dagger\right]_{ab} {\mathcal Z}_{ab} \left[\chi\cdot Q\cdot \chi^\dagger\right]_{bb} \\
\notag
& &+\left[\chi\cdot Q\cdot h\cdot D\cdot \chi^\dagger\right]_{aa} {\mathcal Z}_{ab} \left[\chi\cdot D\cdot \chi^\dagger\right]_{bb} \\
\notag
& &+\left[\chi\cdot D\cdot \chi^\dagger\right]_{aa} {\mathcal Z}_{ab} \left[\chi\cdot Q \cdot h \cdot D\cdot \chi^\dagger\right]_{bb} \\
\notag
& &-\left[\chi\cdot D\cdot h \cdot Q \cdot \chi^\dagger\right]_{ab} {\mathcal Z}_{ab} \left[\chi\cdot D\cdot \chi^\dagger\right]_{ab} \\
\notag
& &+\left[\chi\cdot Q\cdot \chi^\dagger\right]_{ab} {\mathcal Z}_{ab} \left[\chi\cdot Q \cdot h \cdot D\cdot \chi^\dagger\right]_{ab}
\Big) 
\end{eqnarray}
The operations in \refeqn{EAHB} consist of matrix multiplications and summations over the indices $a$ and $b$.  The matrices $h, D,$ and $Q$ are $L\times L$ 
matrices while the matrix $\chi$ is a $P_A \times L$ matrix.  Provided that $P_A = O(L)$, all of the matrix multiplications will require $O(L^3)$ operations.  
Finally, the sum over indices $a$ and $b$ runs from $1$ to $P_A$.  As long as $P_A$ scales linearly with $L$ the sum over $a$ and $b$ can be performed in 
$O(L^2)$ operations.

The value of the gradient can be calculated by differentiating \refeqn{EAHBexpand} with respect to the various parameters.  
\begin{eqnarray}
dE / dA_0 &=& E / A_0 \\
dE / d\chi_{ai} &=& A_0 \sum_b \Big( \trace{h\cdot D}\left[D\cdot \chi^\dagger\right]_{ia} {\mathcal Z}_{ab} \left[\chi\cdot D\cdot \chi^\dagger\right]_{bb} \\
\notag
&& +\trace{h\cdot D}\left[\chi\cdot D\right]_{ai} {\mathcal Z}_{ab} \left[\chi\cdot D\cdot \chi^\dagger\right]_{bb} \\
\notag
&& +\trace{h\cdot D}\left[\chi\cdot D\cdot \chi^\dagger\right]_{bb} {\mathcal Z}_{ba} \left[D\cdot \chi^\dagger\right]_{ia} \\
\notag
&& +\trace{h\cdot D}\left[\chi\cdot D\cdot \chi^\dagger\right]_{bb} {\mathcal Z}_{ba} \left[\chi\cdot D\right]_{ai} \\
\notag
&& +\trace{h\cdot D}\left[D\cdot \chi^\dagger\right]_{ib} {\mathcal Z}_{ab} \left[\chi\cdot Q\cdot \chi^\dagger\right]_{bb} \\
\notag
&& +\trace{h\cdot D}\left[\chi\cdot D\right]_{bi} {\mathcal Z}_{ba} \left[\chi\cdot Q\cdot \chi^\dagger\right]_{aa} \\
\notag
&& +\trace{h\cdot D}\left[\chi\cdot D\cdot \chi^\dagger\right]_{ba} {\mathcal Z}_{ba} \left[Q\cdot \chi^\dagger\right]_{ia} \\
\notag
&& +\trace{h\cdot D}\left[\chi\cdot D\cdot \chi^\dagger\right]_{ba} {\mathcal Z}_{ba} \left[\chi\cdot Q\right]_{ai} \\
\notag
&& +\left[Q\cdot h\cdot D\cdot \chi^\dagger\right]_{ia} {\mathcal Z}_{ab} \left[\chi\cdot D\cdot \chi^\dagger\right]_{bb} \\
\notag
&& +\left[\chi\cdot Q\cdot h\cdot D\right]_{ai} {\mathcal Z}_{ab} \left[\chi\cdot D\cdot \chi^\dagger\right]_{bb} \\
\notag
&& +\left[\chi\cdot Q\cdot h\cdot D\cdot \chi^\dagger\right]_{bb} {\mathcal Z}_{ba} \left[D\cdot \chi^\dagger\right]_{ia} \\
\notag
&& +\left[\chi\cdot Q\cdot h\cdot D\cdot \chi^\dagger\right]_{bb} {\mathcal Z}_{ba} \left[\chi\cdot D\right]_{ai} \\
\notag
&& +\left[D\cdot \chi^\dagger\right]_{ia} {\mathcal Z}_{ab} \left[\chi\cdot Q \cdot h \cdot D\cdot \chi^\dagger\right]_{bb} \\
\notag
&& +\left[\chi\cdot D\right]_{ai} {\mathcal Z}_{ab} \left[\chi\cdot Q \cdot h \cdot D\cdot \chi^\dagger\right]_{bb} \\
\notag
&& +\left[\chi\cdot D\cdot \chi^\dagger\right]_{bb} {\mathcal Z}_{ba} \left[Q \cdot h \cdot D\cdot \chi^\dagger\right]_{ia} \\
\notag
&& +\left[\chi\cdot D\cdot \chi^\dagger\right]_{bb} {\mathcal Z}_{ba} \left[\chi\cdot Q \cdot h \cdot D\right]_{ai} \\
\notag
&& -\left[D\cdot h \cdot Q \cdot \chi^\dagger\right]_{ib} {\mathcal Z}_{ab} \left[\chi\cdot D\cdot \chi^\dagger\right]_{ab} \\
\notag
&& -\left[\chi\cdot D\cdot h \cdot Q\right]_{bi} {\mathcal Z}_{ba} \left[\chi\cdot D\cdot \chi^\dagger\right]_{ba} \\
\notag
&& -\left[\chi\cdot D\cdot h \cdot Q \cdot \chi^\dagger\right]_{ab} {\mathcal Z}_{ab} \left[D\cdot \chi^\dagger\right]_{ib} \\
\notag
&& -\left[\chi\cdot D\cdot h \cdot Q \cdot \chi^\dagger\right]_{ba} {\mathcal Z}_{ba} \left[\chi\cdot D\right]_{bi} \\
\notag
&& +\left[Q\cdot \chi^\dagger\right]_{ib} {\mathcal Z}_{ab} \left[\chi\cdot Q \cdot h \cdot D\cdot \chi^\dagger\right]_{ab} \\
\notag
&& +\left[\chi\cdot Q\right]_{bi} {\mathcal Z}_{ba} \left[\chi\cdot Q \cdot h \cdot D\cdot \chi^\dagger\right]_{ba} \\
\notag
&& +\left[\chi\cdot Q\cdot \chi^\dagger\right]_{ab} {\mathcal Z}_{ab} \left[Q \cdot h \cdot D\cdot \chi^\dagger\right]_{ib}\\
\notag
&& +\left[\chi\cdot Q\cdot \chi^\dagger\right]_{ba} {\mathcal Z}_{ba} \left[\chi\cdot Q \cdot h \cdot D\right]_{bi} \Big) \\
dE/d{\mathcal Z}_{ab} &=& A_0\sum_{ab}\Big(\trace{h\cdot D}\left[\chi\cdot D\cdot \chi^\dagger\right]_{aa} \left[\chi\cdot D\cdot \chi^\dagger\right]_{bb} \\
\notag
&&+\trace{h\cdot D}\left[\chi\cdot D\cdot \chi^\dagger\right]_{ab} \left[\chi\cdot Q\cdot \chi^\dagger\right]_{bb} \\
\notag
&&+\left[\chi\cdot Q\cdot h\cdot D\cdot \chi^\dagger\right]_{aa} \left[\chi\cdot D\cdot \chi^\dagger\right]_{bb} \\
\notag
&&+\left[\chi\cdot D\cdot \chi^\dagger\right]_{aa} \left[\chi\cdot Q \cdot h \cdot D\cdot \chi^\dagger\right]_{bb} \\
\notag
&&-\left[\chi\cdot D\cdot h \cdot Q \cdot \chi^\dagger\right]_{ab} \left[\chi\cdot D\cdot \chi^\dagger\right]_{ab} \\
\notag
&&+\left[\chi\cdot Q\cdot \chi^\dagger\right]_{ab} \left[\chi\cdot Q \cdot h \cdot D\cdot \chi^\dagger\right]_{ab}\Big) 
\end{eqnarray}
Notice that many of the quantities involved in the above equations are repeated.  These can be calculated and stored for later use, to speed computation
at the expense of a larger memory requirement.  The expressions above provide a representative example of the kinds of calculations involved in our 
computation.  


\end{document}